\documentstyle[12pt]{article}
\newcommand{\be}{\begin{equation}}
\newcommand{\bea}{\begin{equarray}}
\newcommand{\ee}{\end{equation}}
\newcommand{\eea}{\end{equarray}}

\begin{document}
\parskip=6pt
\baselineskip=20pt
\smallskip
\bigskip
\smallskip

\bigskip

\centerline{\large \bf Solutions of n-simplex Equation from Solutions }
\bigskip
\centerline {\large \bf of Braid Group Representation
\footnote{This research was partially supported by the China center of advanced
science
and technology and the NNSF of China.}}
\vspace{4ex}
\smallskip
\centerline{~ \sc You-Quan Li$^a$~~~Zhan-Ning Hu$^{b,}$ \footnote{\bf email
address: huzn@itp.ac.cn}}
\smallskip
\centerline{CCAST(World Lab.), P.O.Box 8730, Beijing 100080}
\centerline{$^a$ Zhejiang Institute of Modern Physics, Zhejiang University,
Hangzhou 310027}
\centerline{$^b$ Institute of Theoretical Physics, Academia Sinica,}
\centerline{P. O. Box 2735, Beijing 100080, China \footnote{\bf mail address}}

\vspace{4ex}
\bigskip
\begin{center}
\begin{minipage}{5in}
\centerline{\large\bf 	Abstract}
\vspace{3ex}

It is shown that a kind of solutions of n-simplex equation can be obtained
from representations of braid group. The symmetries in its solution space
are also discussed.

\end{minipage}
\end{center}
\newpage

Recently many interests have been paid on the investigations of the higher
 dimensional integrable systems in the quantum field theory \cite{1} and in
 the statistical mechanics \cite{2}.
 For the lower dimension case of them, the Yang-Baxter equation (YBE) plays a
 crucial role of which the structure is now fairly well understood. As a
 substitution of YBE the tetrahedron equation becomes a integrability
 condition of the exactly solved model in three dimensions \cite{3}, from
which the community of the layer-to-layer transfermatrixes is preserved. One
 of the approaches is the $n$-simplex equation \cite{4} and it is said that
 the case of $n=3$ is corresponding to the tetrahedron equation.
The aim of this letter is to expose some procedure for deriving solutions of
n-simplex equation from  braid group representations ( ie. solutions of
parameter independent  YBE) \cite{5}. Meanwhile we would like to derive
some symmetry transformation in solution space of 3-simplex equation as an
example.

The 3-simplex equation we will consider takes the following form
\be
R_{123}R_{214}R_{341}R_{432} = R_{234}R_{143}R_{412}R_{321}
\ee
where the order of subscripts are chosen in such a way that the normal of
each surface of the 3-simplex is always toward the inside of the 3-simplex
(tetrahedron)
\vskip 2cm

{}~~

Certainly, the  positive direction of the normal of a surface determined by a
cycle ( for example, (123), (341) etc.) following the right-hand helicity. The
matrices in eq.(1) stands for the scattering of three strings, for example
\be
R_{214} | \mu_{1}, \mu_{2}, \mu_{3}, \mu_{4} > =
\sum_{\nu_{1} \nu_{2} \nu_{3} } R^{\nu_{2} \nu_{1} \nu_{4} }_{
\mu_{2} \mu_{1} \mu_{4} } |\nu_{1}, \nu_{2}, \mu_{3}, \nu_{4} >.
\ee
Solving solutions of eq.(1) is a complicated problem. It is  known that many
representations of braid group have been found in recently years. We will
show that if one have a representation of braid group, one can obtain a kind
of solutions of the 3-simplex equation. A braid group is a category of
free group under the constraint of the following equivalence relations
$$
b_{i}b_{i+1}b_{i}  =  b_{i+1}b_{i}b_{i+1}
$$
\be
b_{i}b_{j} =  b_{j}b_{i} ~~~~ for ~~~~|i-j| > 1.
\ee
It is called a braid group due to it has a simple realization on N-strings
by identify

\vskip 2cm
Then the equivalence relation (3) becomes an evident topological equivalence
relation. If a representation of braid group takes as
\be
\rho :  b_{i} \rightarrow S_{i,i+1} = I^{(1)}\otimes\cdots I^{(i-1)}
\otimes S \otimes I^{(i+2)} \otimes \cdots I^{(N)}
\ee
where $ S \in End(V\otimes V) $ satisfying the following parameter independent
Yang-Baxter equation
\be
S_{12}S_{23}S_{12} = S_{23}S_{12}S_{23}.
\ee
If we define an operator
$$
t := \prod_{i=1}^{n}\prod_{j=1}^{i-1} b_{i}
$$
which is understood as an ordered product from right to left or vise versa.
We can show that the following identity holds
\be
t_{1}t_{2}t_{1}t_{2}\cdots = t_{2}t_{1}t_{2}t_{1}\cdots,
\label{eq:z}
\ee
where the number of $t'$s in alternative product is $n+1$.
The case $ n=2$ is exactly the elementary equivalence relations of braid group
eq.(3). For $n=3$ we have
\be
t_{1}t_{2}t_{1}t_{2} = t_{2}t_{1}t_{2}t_{1},
\ee
where $t_{1} = b_{1}b_{2}b_{1},~~ t_{2} = b_{2}b_{3}b_{2}$. Thus if we know a
representation of braid  group, we will have a solution of the following
equation
\be
\check{R}_{123}\check{R}_{234}\check{R}_{123}\check{R}_{234} =
\check{R}_{234}\check{R}_{123}\check{R}_{234}\check{R}_{123}
\label{eq:h}
\ee
where $\check{R}_{123} := \check{R} \otimes I, \check{R}_{234} :=
I \otimes \check{R}$ and $\check{R} \in End(V\otimes V\otimes V)$.
This is easily realized by
$$
\rho : t_{1} \rightarrow \check{R}_{123}
$$
due to $t_{1} = b_{1}b_{2}b_{1} $ etc., then the following identities holds
\be
\check{R}_{123} = S_{12}S_{23}S_{12}\otimes I ~~~etc.
\ee
As to eq.(\ref{eq:h}), one may find some symmetry transformation of it. If one
write out eq.(\ref{eq:h}) into component form instead of matrix form, one can
easily find that the equation can be symbolized by Kauffman diagram. That
says if we denote

\vskip 0.3cm
$$
\check{R}^{abc}_{def} ~~~~~~~~~~~~~~~,~~~
\check{R^{-1}}^{abc}_{def} ~~~~~~~~~~~~~~~
$$
\vskip 0.2cm
The inverse relation and eq.(\ref{eq:h}) are depicted respectively as

\vskip 2cm
and
\vskip 4cm
\be
\label{eq:a}
\ee
where the inner line connecting legs of two shadows implies the
summation over the repeated labels on the legs, and a simple vertical
line stands for a unit matrix. It is not difficult to find that the diagram
eq.(\ref{eq:a}) has the following symmetries:

Flipping via a horizontal axis, denoted by H
\vskip 2.5cm

\be
\label{eq:b}
\ee
or flipping via a vertical axis denoted by V
\vskip 2.5cm

\be
\label{eq:c}
\ee
or via both in term $VH=HV$.
\vskip 2.5cm

\be
\label{eq:d}
\ee

Thus we have
$$
\begin{array}{ccc}
(7)& \stackrel{H}{ \rightarrow} & (8) \cr
V \downarrow &     ~ & \downarrow V\cr
(9) &\stackrel{H}{ \rightarrow} & (10)
\end{array}
$$
and $H^{2} = id, V^{2} = id, HV=VH.$ All the four diagrams eq.(\ref{eq:a}),
eq.(\ref{eq:b}), eq.(\ref{eq:c}), eq.(\ref{eq:d}) depict the same equation
eq.(\ref{eq:h}).
So the solution space
of eq.(\ref{eq:h}) has a discrete group symmetry.
$\{ id, H, V, VH | H^{2} = id, V^{2} = id, HV=VH \}. $ The action of this
group brings one solution of ) into other three new solutions.
i.e. if $\check{R}^{abc}_{def}$ is a solution of eq.(\ref{eq:h}), then
$\check{R'}^{abc}_{def} = \check{R}^{cba}_{fed}$,
$\check{R''}^{abc}_{def}= \check{R}^{edf}_{abc}$ and
$\check{R'''}^{abc}_{def} = \check{R}^{fed}_{cba}$ will be solutions of
eq.(\ref{eq:h}).

Furthermore, if giving a direction to the Kauffnman diagram
$\check{R}^{abc}_{def}$
$~~~~~~~~~~~~~~$, and adding a minus sign to the labels on the tip
of the arrow,
we can find that the summation of such labels on both side of the diagram
eq.(\ref{eq:a})
are equal. This brings about a contineous transformation from a solution of
eq.(\ref{eq:h}) into another
\be
\check{R}^{abc}_{def} \rightarrow \check{R'}^{abc}_{def}
= t^{a+b+c-d-e-f}\check{R}^{abc}_{def}.
\ee
Starting from the matrix form of eq.(\ref{eq:h}), we can obtain two more
contineous
transformations in solution space. They are an overall factor transformation
$ \check{R}\rightarrow \tau \check{R} $; a similar transformation by a tensor
product of matrices $\check{R} \rightarrow (\Lambda \otimes \Lambda \otimes
\Lambda ) \check{R} (\Lambda^{-1} \otimes \Lambda^{-1} \otimes \Lambda^{-1})$.
Because  eigenvalues of a matrix are invariant under a similar
transformation, the latter is a transformation within the subset of solution
space, which is specialfied by the eigenvalues of $\check{R}$.

In above we made much discussion on eq.(\ref{eq:h}), now we introduce a new
$R$-matrix
\be
\check{R} = R P
\ee
where $ P$ is defined as
$$
P| \mu_{1}, \mu_{2}, \mu_{3} > := | \mu_{3}, \mu_{2}, \mu_{1} >.
$$
Then we can show the $R$-matrix satisfing the following equation as
long as the $\check{R}$-matrix satisfing eq.(\ref{eq:h})
 \be
R_{123}R_{214}R_{341}R_{432} = R_{234}R_{143}R_{412}R_{321}
\ee
which is an variant of the FM 3-simplex equation we have introduced at the
begining of our discussion.

In a similar way, one may discuss the case of 4-simplex equation and so on.
The key point is that eq.(\ref{eq:z}) is an identity on braid group, then if
 one has
a representation of braid group, one can write down a expression from
the expression of $t_{i}$ on the basis of $S$-matrix, which is supposed
to be solutions of parameter independent Yang-Baxter equation.

\centerline{$\ast \ast \ast$}
One of the authors (Hu) would like to thank H. Yan for the interesting
 discussions.

\end{document}